\documentclass[article]{article}
\usepackage{color}
\usepackage{graphicx}
\usepackage{amsmath}
\usepackage{amsmath,amssymb,amsthm,amsfonts,mathrsfs}
\usepackage{amssymb}
\usepackage{amsthm}
\usepackage{mathrsfs}
\usepackage[numbers,sort&compress]{natbib}
\usepackage{amsthm}
\usepackage{amsbsy}
\usepackage{caption}
\usepackage{pgf}
\usepackage{CJK}
\usepackage{ctex}
\usepackage{graphicx}
\usepackage{tikz}
\usepackage{stmaryrd}
\usepackage{graphicx}
\usepackage{hyperref}
\usepackage[latin1]{inputenc}
\usepackage[T1]{fontenc}
\usepackage[english]{babel}
\usepackage{listings}
\usepackage{xcolor,mathrsfs,url}
\usepackage{amssymb}
\usepackage{amsmath}
\usepackage{ifthen}

\begin{document}

\title{ Inverse scattering transformation  for the Fokas-Lenells equation with nonzero boundary conditions  }
\author{Yi ZHAO$^1$ and Engui FAN$^{1}$\thanks{\ Corresponding author and email address: faneg@fudan.edu.cn } }
\footnotetext[1]{ \  School of Mathematical Sciences, Fudan University, Shanghai 200433, P.R. China. }

\date{ }

\maketitle
\begin{abstract}
\baselineskip=17pt
In this article, we focus on the inverse scattering transformation  for the Fokas-Lenells (FL) equation with nonzero boundary conditions 
 via the Riemann-Hilbert (RH) approach. Based on the Lax pair of the FL equation, the analyticity and  symmetry, asymptotic behavior of Jost solutions and scattering matrix  are discussed in detail.
 With these results, we further present a generalized RH problem, from
 which  a  reconstruction formula between the solution of  the FL equation  and  the Riemann-Hilbert  problem
 is obtained.    The  N-soliton solutions of the FL equation   is obtained via solving the RH problem.
\end{abstract}

\baselineskip=17pt

\newpage

\section{Introduction}

 The nonlinear schr\"{o}dinger(NLS) equation
  \begin{equation}
i u_t+u_{xx}-2\nu|q|^2q=0 \label{NLS}
\end{equation}
($\nu=-1$ and $\nu=1$ denote the focusing and defocusing cases), is a significant mathematical and physical model for the optical fibers, deep water waves and plasma physics  \cite{biondini2014inverse}.   The NLS equation is one of most integrable systems, which  admits  many special features,  including the existence of a Lax pair and the possession of a bi-Hamiltonian formulation \cite{shabat1972exact, fokas2012important}.

 In the late seventies,   standard bi-Hamiltonian formulation was used to   obtain   an integrable generalization of a given equation   \cite{fuchssteiner1981symplectic}.   For instance, the Cammssa-Holm equation is derived  from the KdV equation via  the bi-Hamiltonian structure \cite{Fokas1995},  and the same mathematical trick can be applied to the NLS equation yields
 Fokas-Lenells  (FL) equation  \cite{lenells2008novel}
\begin{equation}
iu_t-\nu u_{tx}+\gamma u_{xx}+\sigma |u|^2(u+i\nu u_x)=0,\quad \sigma=\pm 1.
\label{2}
\end{equation}
If let $\alpha=\gamma/\nu>0$ and $\beta=\frac{1}{\nu}$ and
$$u\rightarrow \beta\sqrt\alpha e^{i\beta x} u,\quad \sigma\rightarrow-\sigma,$$
then equation (\ref{2}) can be  converted   into
\begin{equation}
u_{tx}+\alpha\beta^2u-2i\alpha\beta u_x-\alpha u_{xx}+\sigma i\alpha\beta^2|u|^2u_x=0,
\label{FL1}
\end{equation}
where $\alpha, \ \beta>0$,   and here  we   consider the focusing case of $\sigma=-1$.

 In recent years, much work has been done on the FL equation (\ref{FL1}).   For example,    Lax pair for FL equation was
 obtained  via   the bi-Hamiltonian structure by  Fokas and Lenells  \cite{lenells2008novel}.
 They further considered  the initial-boundary problem for the FL  equation  (\ref{2})   on the half-line by using Fokas unified method \cite{lenells2009integrable};   The    dressing method is applied to obtain an  explicit formula for N-bright-soliton solutions for the equation (\ref{2}) \cite{lenells2010dressing};    The N-dark soliton solution is obtained in \cite{vekslerchik2011lattice}.
  The bilinear method was used to obtain   bright and dark $N$-soliton solution are obtained via a systematic method \cite{Matsuno_2012,matsuno2012direct}.
   The Darboux transformation is used to
 obtain   rogue waves of the FL equation  \cite{he2012rogue,xu2015n}.  The Deift-Zhou nonlinear steepest decedent method was used
 to analyze long-time  asymptotic behavior  for FL equation with decaying initial value \cite{xu2015}.      Riemann-Hilbert approach
 was adopted to construct    explicit  soliton solutions  under zero boundary conditons\cite{ai2019riemann}.
 As far as we know,  the soliton solutions  for  the FL equation  (\ref{FL1}) with nonzero boundary conditions  (NZBCs)  have not been reported.
  In this paper, we  apply Riemann-Hilbert (RH) approach to focus on the inverse transformation  of   the FL equation (\ref{FL1} with the following NZBCs
\begin{equation}
u(x,t)\sim q_{\pm}e^{-i\beta x+2i\alpha\beta  t},\quad x\rightarrow\pm\infty,
\end{equation}
where $|q_{\pm}|= \sqrt{\frac{2}{\beta}}$.

\indent The inverse scattering transform is an important method to study   physically important nonlinear wave equations with Lax pair such as the NLS equation,  the modified KdV equation, Sine-Gordan equation  \cite{novikov1984theory,beals1984scattering}.    As an improved version of inverse scattering transform, the Riemann-Hilbert method has been widely adopted to solve nonlinear integrable models \cite{biondini2016inverse,pichler2017focusing,WF20191,ZF2019,WF20192,YF2019}

\indent  The paper is organized as follows. In Section 2, starting from a Lax pair, a transformation is introduced to invert the boundary conditions into constants,  and then Jost solutions are obtained.  Furthermore, the analyticity, symmetry and asymptotic behavior  associated with   eigenfunctions and  scattering matrix are obtained.   In Section 3,  a generalized
RH  problem for  the FL equation is constructed,  and the distribution of discrete spectrum and residue conditions associated with RH problem  are discussed.   Based on these results, we reconstruct the potential function.  In Section 4,  we gives  the N-soliton solutions  via solving RH problem under reflectionless case.

\section{The direct scattering with NZBCs}

\subsection{Jost solutions}

\noindent It is well-known that the FL equation (\ref{FL1}) admits a Lax pair
\begin{equation}
\begin{split}
&\psi_x+ik^2\sigma_3\psi=kU_x\psi,\\
&\psi_t+i\eta^2\sigma_3\psi=\left[\alpha k U_x+\frac{i\alpha\beta^2}{2}\sigma_3\left(\frac{1}{k}U-U^2\right)\right]\psi,
\end{split}
\label{lax1}
\end{equation}
where
\begin{equation}
U=\left(\begin{array}{cc}
        0& u\\
        v&0
        \end{array}
  \right),\quad \sigma_3=\left(\begin{array}{cc}
        1& 0\\
        0&-1
        \end{array}
  \right),\quad \eta=\sqrt\alpha\left(k-\frac{\beta}{2k}\right),\quad v=-u^*.
\nonumber
\end{equation}

\noindent In order to invert the nonzero boundary conditions into the constant boundary conditions,   we   introduce a transformation for eigenfunctions  and potentials.\\
\noindent\textbf{Theorem 2.1 } By transformation
\begin{equation}
\begin{split}
&u=qe^{-i\beta x+2i\alpha\beta t},\\
&\psi=e^{\left(-\frac12i\beta x+ i\alpha\beta t\right)\sigma_3}\varphi,
\label{qu}
\end{split}
\end{equation}
the FL equation (\ref{FL1}) becomes
\begin{equation}
q_{tx}-i\beta q_t+2i\alpha\beta q_x-\alpha q_{xx}+(2\alpha\beta^2 -\alpha\beta^3|q|^2)q-i\alpha\beta^2|q|^2q_x=0
\label{FL2}
\end{equation}
with corresponding boundary conditions
\begin{equation}
q\rightarrow q_{\pm}, \quad x\rightarrow\pm\infty.
\end{equation}
And Eq. (\ref{FL2}) is the compatibility condition of the Lax pair
\begin{equation}
\varphi_x=X\phi,\quad \varphi_t=T\varphi,
\label{lax2}
\end{equation}
where
\begin{equation}
\begin{split}
X&=-ik^2\sigma_3+\frac12i\beta\sigma_3-i\beta k\sigma_3Q+kQ_x,\\
T&=-i\eta^2\sigma_3-\frac12i\alpha\beta^2\sigma_3Q^2-\frac12i\alpha\beta^2q_0^2\sigma_3\\
&-i\alpha\beta k\sigma_3 Q+\alpha k Q_x+\frac{i\alpha\beta^2}{2k}\sigma_3Q.
\end{split}
\nonumber
\end{equation}
\begin{proof}
Suppose the transformation is
\begin{equation}
\begin{split}
&u=qe^{Aix+Bit},\ \ \psi=e^{(iCx+iDt)\sigma_3}\varphi.
\end{split}
\nonumber
\end{equation}
Substituting it into (\ref{FL1}) yields
\begin{equation}
\begin{split}
&q_{xt}+Aiq_t+i(B-2\alpha\beta-2\alpha A)q_x+(2\alpha\beta A+\alpha\beta^2+\alpha A^2-AB)q\\
-&\alpha q_{xx}+A\alpha\beta^2|q|^2q-i\alpha\beta^2|q|^2q_x=0.
\end{split}
\end{equation}
And the Lax pair becomes
\begin{equation}
\begin{split}
\varphi_x=&e^{\left[i\left(\frac A2-C\right)x+i\left(\frac B 2-D\right)t\right]\hat{\sigma_3}}
 \left(\begin{array}{cc}
      -ik^2-iC & k(Aiq+q_x)\\
      k(Aiq^*-q_x^*) & ik^2+iC
      \end{array}\right)\varphi,\\
\varphi_t=&e^{\left[i\left(\frac A2-C\right)x+i\left(\frac B 2-D\right)t\right]\hat{\sigma_3}}
 \left(\begin{array}{cc}
       -i\eta^2+\frac12i\alpha\beta^2|q|^2-iD & \alpha k(Aiq+q_x)+\frac{i\alpha\beta^2}{2k}q\\
       -\alpha k(-Aiq^*+q_x^*)+\frac{i\alpha\beta^2}{2k}q^* & i\eta^2-\frac12i\alpha\beta^2|q|^2+iD
       \end{array}\right)\varphi.
\end{split}
\end{equation}
Take $A=2C$ and $B=2D$ in the above equation and consider the limit as $x\rightarrow\pm\infty$, one gets
\begin{equation}
\begin{split}
X_{\pm}=&\left(\begin{array}{cc}
               -ik^2-\frac A2i & kAiq_{\pm}\\
               kAiq_{\pm}^* & ik^2+\frac{A}{2}i
               \end{array}
\right),\quad\\
T_{\pm}=&\left(\begin{array}{cc}
            -i\eta^2+ i\alpha\beta -iD    & \alpha k iA q_{\pm}+\frac{i\alpha\beta^2}{2k}q_{\pm}\\
            \alpha k i A q_{\pm}^*+\frac{i\alpha\beta^2}{2k}q_{\pm}^*    & i\eta^2- i\alpha\beta +iD
            \end{array}
\right).
\end{split}\nonumber
\end{equation}
 These two matrices are proportional if
 $$A=-\beta, \quad B=2\alpha\beta.$$
Moreover,  $T_{\pm}$ and $X_{\pm}$ are proportional
$$T_{\pm}=\frac{\sqrt \alpha\eta}{k}X_{\pm}.$$
which   implies that  their eigenvalues are proportional and they share the same eigenvector matrices.
\end{proof}

To diagonalize the matrices $X_{\pm}$ and $T_{\pm}$ and further obtain the Jost solutions, we need to get the eigenvalues and eigenvector matrices of them. Direct calculation shows that the eigenvalues of the matrices $X_{\pm}$
are $\pm\frac{ik}{\sqrt{\alpha}}\lambda$;  and the eigenvalues of the matrices $T_{\pm}$ are $\pm i\eta\lambda$,
where $$\lambda=\sqrt{\alpha}\big(k+\frac{\beta}{2k}\big),$$ and the corresponding eigenvector matrices are
$$Y_{\pm}(k)={\rm I}-\frac{\beta}{2k}Q_{\pm}, \ \ \det{Y_{\pm}(k)}=1+\frac{\beta}{2k^2}\triangleq\gamma(k). $$

\indent With the above results, the matrices can be diagonalized as:
\begin{equation}
\begin{split}
&X_{\pm}=Y_{\pm}\left(-\frac{ik\lambda\sigma_3}{\sqrt\alpha}\right)Y_{\pm}^{-1},\\
&T_{\pm}=Y_{\pm}(-i\eta\sigma_3)Y_{\pm}^{-1}.
\end{split}
\label{diag}
\end{equation}
Thus,  the asymptotic Lax pair
\begin{equation}\
\widetilde{\varphi}_x=X_{\pm}\widetilde\varphi,\quad \widetilde{\varphi}_t=T_{\pm}\widetilde\varphi
\label{aymlax}
\end{equation}
can be written as
$$(Y_{\pm}^{-1}\widetilde\varphi)_x=-\frac{ik\lambda\sigma_3}{\sqrt\alpha}(Y_{\pm}^{-1}\widetilde\varphi),
\ \ (Y_{\pm}^{-1}\widetilde\varphi)_t=-i\eta\sigma_3(Y_{\pm}^{-1}\widetilde\varphi), $$
which have a solution
\begin{equation}
\widetilde\varphi=Y_{\pm}(k)e^{-i\theta(x,t,k)\sigma_3}.
\end{equation}
Therefore, the  asymptotic of the eigenfunctions $\varphi_{\pm}$ are
\begin{equation}
\varphi_{\pm}\sim Y_{\pm}(k)e^{-i\theta(x,t,k)\sigma_3},\quad x\rightarrow\pm\infty,
\end{equation}
where $$\theta(x,t,k)=\frac{k\lambda x}{\sqrt\alpha}+\lambda\eta t.$$
\indent Define the Jost solutions as
\begin{equation}
J_{\pm}(x,t,k)=\varphi_{\pm}(x,t,k)e^{i\theta\sigma_3},
\label{tran1}
\end{equation}
then  the Lax pair (\ref{lax2}) is changed to
\begin{equation}
\begin{split}
&J_{\pm,x}-\frac{ik\lambda}{\sqrt\alpha}J_{\pm}\sigma_3=XJ_{\pm},\\
&J_{\pm,t}-i\lambda\eta J_{\pm}\sigma_3=TJ_{\pm},
\label{18}
\end{split}
\end{equation}
and
\begin{equation}
J_{\pm}\sim Y_{\pm}(k),\quad x\rightarrow\pm\infty.
\end{equation}
\subsection{Scattering matrix}
\noindent The functions $\varphi_{\pm}$ are both fundamental matrix solution of the
Lax pair (\ref{lax2}), thus there exists a matrix $S(k)$ that only depends on $k$, such that
\begin{equation}
\varphi_+(x,t,k)=\varphi_-(x,t,k)S(k),
\label{scamatrix}
\end{equation}
where  $S(k)$ is called scattering matrix. Columnwise, Eq. (\ref{scamatrix}) reads
\begin{equation}
\varphi_{+,1}=s_{11}\varphi_{-,1}+s_{21}\varphi_{-,2},\quad \varphi_{+,2}=s_{12}\varphi_{-,1}+s_{22}\varphi_{-,2}.\label{linear}
\end{equation}
\indent Since $tr(X)=tr(T)=0$, according to the Abel's formula, we have
$$({\rm det}\ \varphi_{\pm})_x=({\rm det}\ \varphi_{\pm})_t=0,$$
which implies
$${\rm det}\ \varphi_{\pm}=\lim_{x\rightarrow\pm\infty}{\rm det}\ \varphi_{\pm}=\gamma(k),$$
then the scattering coefficients can be expressed as Wronskians of columns $\varphi_{\pm}$ in the following way
\begin{equation}
\begin{split}
&s_{11}(k)=\frac{Wr(\varphi_{+,1},\varphi_{-,2})}{\gamma(k)},\quad s_{12}(k)=\frac{Wr(\varphi_{+,2},\varphi_{-,2})}{\gamma(k)},\\
&s_{21}(k)=\frac{Wr(\varphi_{-,1},\varphi_{+,1})}{\gamma(k)},\quad s_{22}(k)=\frac{Wr(\varphi_{-,1},\varphi_{+,2})}{\gamma(k)}.
\end{split}\label{wrons}
\end{equation}
\subsection{Asymptotic analysis}
\noindent To properly construct the Riemann-Hilbert problem, we need consider the asymptotic
behavior of eigenfunctions and scattering matrix as  $k\rightarrow\infty$ and $k\rightarrow 0$.

\subsection{Asymptotic as $k\rightarrow\infty$}
Consider a solution of (\ref{18}) of the form
$$J_{\pm}=J_{\pm}^{(0)}+\frac{J_{\pm}^{(1)}}{k}+\frac{J_{\pm}^{(2)}}{k^2}+\cdots,$$
then substituting the above expansion into (\ref{18}) and comparing the coefficients of the same order of $k$, we get the comparison results:\\
x-part:
\begin{align}
O(k^0): J_{\pm,x}^{(0)}+i[\sigma_3,J_{\pm}^{(2)}]&=i\beta \sigma_3J_{\pm}^{(0)}-i\beta\sigma_3QJ_{\pm}^{(1)}+Q_xJ_{\pm}^{(1)},\\
O(k^2):J_{\pm}^{(0)}\sigma_3&=\sigma_3J_{\pm}^{(0)}.
\end{align}
t-part:
\begin{align}
O(k):[\sigma_3,J_{\pm}^{(1)}]=&-\beta\sigma_3QJ_{\pm}^{(0)}-iQ_xJ_{\pm}^{(0)},\label{21}\\
O(k^0):J_{\pm,t}^{(0)}+i\alpha[\sigma_3,J_{\pm}^{(2)}]=&i\alpha\beta\sigma_3J_{\pm}^{(0)}-\frac12i\alpha\beta^2Q^2\sigma_3J_{\pm}^{(0)}-\frac12i\alpha\beta^2q_0^2\sigma_3J_{\pm}^{(0)}\nonumber\\
&-i\alpha\beta\sigma_3QJ_{\pm}^{(1)}+\alpha Q_xJ_{\pm}^{(1)}.
\end{align}
\indent Based on these results, we derive that $J_{\pm}^{(0)}$ is diagonal and satisfies
\begin{align}
J^{(0)}_{\pm,x}=&i\nu_1\sigma_3J_{\pm}^{(0)},\label{23}\\
J^{(0)}_{\pm,t}=&i\nu_2\sigma_3J_{\pm}^{(0)}\label{24},
\end{align}
where
\begin{equation}
\begin{split}
\nu_1(x,t)=&\beta+\frac12\beta^2qr-\frac12i\beta qr_x+\frac12i\beta q_xr+\frac12q_xr_x,\\
\nu_2(x,t)=&\alpha\beta-\frac12\alpha\beta^2q_0^2+\frac12i\alpha\beta(q_xr-qr_x)+\frac12\alpha q_xr_x  .
\end{split}\nonumber
\end{equation}
Note that the FL equation (\ref{FL2}) admits the conservation law
$$\left( q_xr_x-i\beta qr_x+i\beta q_xr+\beta^2qr \right)_t=\left( \alpha q_xr_x+i\alpha\beta(q_xr-qr_x) \right)_x,$$
thus equations  (\ref{23}) and   (\ref{24})  are consistent and are both satisfied if we define
\begin{equation}
J_{\pm}^{(0)}=e^{i\nu\sigma_3},\quad \nu=\int_{-\infty}^x\left( \beta+\frac12\beta^2qr-\frac12i\beta qr_x+\frac12i\beta q_xr+\frac12q_xr_x  \right)dx^{'}.\label{27}
\end{equation}
Therefore, we obtain the limit of the Jost solutions as $k\rightarrow\infty$:
\begin{equation}
J_{\pm}\sim J_{\pm}^{(0)},\quad k\rightarrow \infty.
\end{equation}
\indent Define
\begin{equation}
J_{\pm}=J_{\pm}^{(0)}\mu_{\pm},
\label{tran2}
\end{equation}
then we have
\begin{equation}
\mu_{\pm}\rightarrow {\rm I},\quad k\rightarrow\infty.
\end{equation}
In addition, the asymptotic property of $\varphi_{\pm}$    follows:
$$\varphi_{\pm}\sim e^{i(\nu-\theta)\sigma_3},\quad k\rightarrow\infty.$$
Consider the wronskians expression of the scattering coefficients (\ref{wrons}), we find that
\begin{equation}
s_{11}(k)\rightarrow 1, \ s_{22}(k)\rightarrow 1,\quad k\rightarrow\infty.
\end{equation}
\subsection{Asymptotic as $k\rightarrow 0$}
Consider a solution of (\ref{18}) of the form
$$J_{\pm}=\frac{D_{\pm}^{(-1)}}{k}+D_{\pm}^{(0)}+D_{\pm}^{(1)}k+\cdots,$$
then   we obtain the comparison results:\\
x-part:
\begin{equation}
O(k^{-1}):D_{\pm,x}^{(-1)}-\frac12i\beta D_{\pm}^{(-1)}\sigma_3=\frac12 i\beta\sigma_3D_{\pm}^{-1},
\end{equation}
t-part:
\begin{equation}
O(k^{-3}):D_{\pm}^{-1}\sigma_3=-\sigma_3D_{\pm}^{-1}.
\end{equation}
It is easy  to check that
\begin{equation}
D_{\pm,x}^{(-1)}=0,
\end{equation}
which  implies $D_{\pm}^{(-1)}$ is a constant independent on $x$. Taking limit on $kJ_{\pm}$   yields
$$\lim_{x\rightarrow\pm\infty}\lim_{k\rightarrow 0} kJ_{\pm}=\lim_{k\rightarrow 0}\lim_{x\rightarrow\pm\infty}kJ_{\pm}=-\frac{\beta}{2}Q_{\pm},$$
thus we know $$D_{\pm}^{(-1)}=-\frac{\beta}{2}Q_{\pm},$$ substituting into (\ref{18}) leads to
\begin{equation}
J_{\pm}=-\frac{\beta Q_{\pm}}{2k}+O(1),\quad k\rightarrow0,
\end{equation}
then the asymptotic behaviour of functions $\varphi_{\pm}$ and $\mu_{\pm}$ follows. Consider the wronskian   expressions of scattering  coefficients and note the boundary conditions, we find that
\begin{equation}
\begin{split}
&s_{11}=Wr(J_{+,1}e^{-i\theta}, J_{-,2}e^{i\theta}) =\frac{q_-}{\beta q_+}+O(1),\quad k\rightarrow 0, \\ &s_{22}=Wr(J_{-,1}e^{-i\theta}, J_{+,2}e^{i\theta}) =\frac{q_-^*}{\beta q_+^*}+O(1),\quad k\rightarrow 0.
\end{split}\nonumber
\end{equation}

\subsection{Analyticity}
Under the transformation (\ref{tran2}), we write the  Lax pair as
\begin{equation}
\begin{split}
\mu_{\pm,x}-\frac{ik\lambda}{\sqrt \alpha}[\mu_{\pm},\sigma_3]=&\hat{X}\mu_{\pm},\\
\mu_{\pm,t}-i\lambda\eta[\mu_{\pm},\sigma_3]=&\hat{T}\mu_{\pm},
\end{split}\label{29}
\end{equation}
where
$$\hat{X}=e^{-i\nu\sigma_3}(X+\frac{ik\lambda\sigma_3}{\sqrt \alpha}-i\nu_1\sigma_3),$$
$$\hat{T}=e^{-i\nu\sigma_3}(T+i\lambda\eta\sigma_3-i\nu_2\sigma_3).$$
Then we obtain a full derivative
\begin{equation}
d(e^{i\theta\hat{\sigma}_3}\mu_{\pm})=e^{i\theta\hat{\sigma}_3}(\hat{X}dx+\hat{T}dt)\mu_{\pm},
\end{equation}
which implies the solution of Eq. (\ref{29}) satisfies the following integral equation:
\begin{align}
&\mu_-=\mu_-^0+\int_{-\infty}^xe^{-\frac{ik\lambda}{\sqrt\alpha}(x-y)\hat{\sigma}_3}\hat{X}(y,t,k)\mu_-(y,t,k)dy,\\
&\mu_+=\mu_+^0-\int_x^{+\infty}e^{-\frac{ik\lambda}{\sqrt\alpha}(x-y)\hat{\sigma}_3}\hat{X}(y,t,k)\mu_+(y,t,k)dy,
\end{align}
where
$$\mu_-^0=Y_-,\quad \mu_+^0=e^{-i\int_{-\infty}^{+\infty}\nu_1(x^{'})dx^{'}\sigma_3}Y_+.$$
Thus, the eigenfunctions can be analytically extended in the complex $k$-plane into the following regions:
\begin{equation}
\mu_{-,1},\mu_{+,2}:\quad D^+,\quad \mu_{+,1},\mu_{-,2}:\quad D^-,
\end{equation}
where
$$D_+=\left\{k\ |\ arg\ k\in \left(0,\frac{\pi}{2}\right)\cup\left(\pi,\frac32\pi\right)\right\},$$
$$D_-=\left\{k\ |\ arg\ k\in \left(\frac{\pi}{2},\pi\right)\cup\left(\frac32\pi,2\pi\right)\right\},$$
which are shown in Fig.1,
and the subscripts 1 and 2 identify matrix columns, i.e., $\mu_{\pm}=(\mu_{\pm,1,\mu_{\pm,2}})$.\\
\indent Apparently, the functions $\varphi_{\pm}$ and $\mu_{\pm}$ share the same analyticity, hence from the wronskians expression of the scattering coefficients, we know that $s_{11}$ is analytic in $D^-$, and $s_{22}$ is analytic in $D^+$.
\begin{center}
\begin{tikzpicture}
\draw [<-](-2.8,0)--(-2.2,0);
\draw [-](-4,0)--(-2.8,0);
\draw [->](-1.4,0)--(-0.6,0);
\draw [->](-2.2,0)--(-1.4,0);

\draw [->](-2.2,-1.5)--(-2.2,-0.7);
\draw [-](-2.2,-0.7)--(-2.2,0);
\draw [->](-2.2,1.5)--(-2.2,0.7);
\draw [-](-2.2,0.7)--(-2.2,0);
\draw [->](-2.2,0.7)--(-2.2,1.5);

\node at (-1.5, 0.7 )  {$  +$};
\node at (-1.5, -0.7 )  {$  -$};
\node at (-2.9, 0.7 )  {$  -$};
\node at (-2.9, -0.7 )  {$  +$};
\node [thick] [above]  at (-2.2,1.5){\footnotesize ${\rm Im}k$};
\node [thick] [above]  at (-0.3,-0.2){\footnotesize ${\rm Re}k$};
\end{tikzpicture}
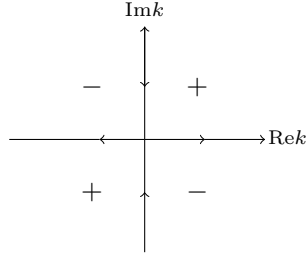
 \center{\footnotesize {\bf Fig.~1.} The jump contour in the complex k-plane}
\end{center}
\subsection{Symmetry}
To investigate the discrete spectrum and residue conditions in the Riemann-Hilbert problem, one needs to analyze the  symmetric property for the solutions $\varphi_{\pm}$ and the scattering matrix $S(k)$. \\
\noindent\textbf{Theorem 2.2. } The Jost eigenfunctions satisfy the following symmetric  relations
\begin{align}
\sigma\varphi_{\pm}^*(k^*)\sigma&=-\varphi_{\pm}(k),\label{34}\\
\sigma_1\varphi_{\pm}^*(-k^*)\sigma_1&=\varphi_{\pm}(k),\label{35}
\end{align}
and the scattering matrix satisfies
\begin{align}
-\sigma S(k)\sigma&=S^*(k^*),\label{36}\\
\sigma_1S^*(-k^*)\sigma_1&=S(k).\label{37}
\end{align}
where
$$\sigma=\left(\begin{array}{cc}
               0 & 1\\
               -1 & 0
               \end{array}
         \right),\quad \sigma_1=\left(\begin{array}{cc}
               0 & 1\\
               1 & 0
               \end{array}
         \right).
$$
\begin{proof}
We only prove  (\ref{34}) and (\ref{36}),   (\ref{35}) and (\ref{37}) can be shown in a similar way.  The functions $\varphi_{\pm}$ are the solutions of the spectral problem
\begin{equation}
\varphi_{\pm,x}(k)=X(k)\varphi_{\pm}(k).
\label{sss}
\end{equation}
Conjugating and multiplying left by $\sigma$ on both sides of the equation gives
$$\left(\sigma\varphi_{\pm}^*(k^*)\sigma\right)_x=\sigma X^*(k^*)\varphi_{\pm}^*(k^*)\sigma.$$
Note that $\sigma X^*(k^*)=X(k)\sigma$, hence $\sigma\varphi_{\pm}^*(k^*)\sigma$ is also a solution of (\ref{sss}).
  By using  relations
$\sigma Y_{\pm}^*(k^*)=Y_{\pm}(k)\sigma$ and $\sigma e^{i\theta\sigma_3}\sigma=-e^{-i\theta\sigma_3}$,  we obtain
\begin{equation}
\sigma \varphi_{\pm}^*(k^*)\sigma\sim-Y_{\pm}(k)e^{-i\theta\sigma_3},\quad x\rightarrow\pm\infty,
\end{equation}
which leads to    (\ref{34}).

For the scattering matrix, conjugating on the both sides of  equation (\ref{scamatrix}) leads to
$$\varphi_{+}^*(k^*)=\varphi_{-}^*(k^*)S^*(k^*),$$
substituting (\ref{34}) into the above formula yields
$$\varphi_{+}(k)=-\varphi_-(k)\sigma S^*(k^*)\sigma,$$
comparing with (\ref{scamatrix}) gives
$$S^*(k^*)=-\sigma S(k)\sigma.$$
\end{proof}
Elementwise, equations (\ref{34})-(\ref{37}) read as
\begin{equation}
\begin{split}
s_{11}^*(k^*)&=s_{22}(k), \quad s_{21}^*(k^*)=-s_{12}(k),\\
s_{11}^*(-k^*)&=s_{22}(k),\quad s_{21}^*(-k^*)=s_{12}(k).
\end{split}\label{40}
\end{equation}
\begin{equation}
\begin{split}
\varphi_{\pm,1}(k)&=\sigma\varphi_{\pm,2}^*(k^*),\quad \varphi_{\pm,2}(k)=-\sigma\varphi_{\pm,1}^*(k^*),\\
\varphi_{\pm,1}(k)&=\sigma_1\varphi_{\pm,2}^*(-k^*),\quad \varphi_{\pm,2}(k)=\sigma_1\varphi_{\pm,1}^*(-k^*),
\end{split}
\end{equation}
which implies
\begin{equation}
\varphi_{\pm,2}(k)=-\sigma_3\varphi_{\pm,2}(-k),\quad \varphi_{\pm,1}(k)=\sigma_3\varphi_{\pm,1}(-k),
\label{42}
\end{equation}
note the relation (\ref{tran1}) and (\ref{tran2}), there exists
\begin{equation}
\varphi_{\pm}=e^{i\nu\sigma_3}\mu_{\pm}e^{-i\theta\sigma_3}, \label{muvar}
\end{equation}
thus for the eigenfunctions $\mu_{\pm}$, we derive
\begin{equation}
\mu_{\pm,2}(k)=-\sigma_3\mu_{\pm,2}(-k),\quad \mu_{\pm,1}(k)=\sigma\mu_{\pm,1}(-k).
\end{equation}

\section{The inverse scattering with NZBCs}
\subsection{Generalized Riemann-Hilbert problem}

We define  the two matrices
$$ M^+=\left(\mu_{-,1},\frac{\mu_{+,2}}{s_{22}} \right),\quad k\in D^+,$$
$$ M^-=\left(\frac{\mu_{+,1}}{s_{11}},\mu_{-,2}\right),\quad k\in D^-,$$
which are analytical in $D^+, \ D^-$ respectively, and admit  asymptotic
\begin{equation}
\begin{split}
M^\pm&=I+O(1/k),\quad k\rightarrow \infty,\\
M^\pm&=\frac{1+\beta}{k}e^{-i\nu\sigma_3}\widetilde{Q}_\pm+O(1),\quad k\rightarrow 0, \label{redf}
\end{split}
\end{equation}
where $\widetilde{Q}_-=\left( \begin{array}{cc}
                        0  & -\frac{1}{q_-^*}\\
                        \frac{1}{q_-} &  0
                       \end{array}\right) $.

By using (\ref{muvar}) (\ref{redf}),   we rewrite  (\ref{linear}) and  get a generalized Riemann-Hilbert problem
\begin{align}
&\star\ M(x,t,k)\ is\ meromorphic\ in\ \rm{C}\setminus\Sigma,\\
&\star\ M^+(x,t,k)=M^-(x,t,k)(I-G(x,t,k)), \quad k\in\Sigma,\label{RHP1}\\
&\star\ M(x,t,k)\ satisfies\ residue\ conditions\ at\ zeros\ \{k: s_{11}(k)=s_{22}(k)=0\},\\
&\star\ {M^{\pm}=I+O(1/k), \quad k\rightarrow\infty,}\\
&\star\ M^\pm =\frac{1+\beta}{k}e^{-i\nu\sigma_3}\widetilde{Q}_\pm+O(1),\quad k\rightarrow 0,
\end{align}
where  the  jump matrix
$$G=\left(\begin{array}{cc}
           0 & -e^{-2i\theta}\widetilde{\rho}(k)\\
           e^{2i\theta}\rho(k) & \rho(k)\widetilde{\rho}(k)
           \end{array}
\right), \ \ \rho(k)=\frac{s_{21}}{s_{11}},\quad \widetilde{\rho}(k)=\frac{s_{12}}{s_{22}}.$$
and
$\Sigma={\rm R}\cup i{\rm R}$ denotes the jump contour in Fig.1.

\subsection{Discrete spectrum and residue conditions}

\noindent The discrete spectrum is the set of all values $k\in {\rm C}\setminus\Sigma$ such that eigenfunctions exist in $L^2({\rm R})$.  As usual, these values are the zeros of $s_{11}(k)$ in $D^+$
 and those of $s_{22}(k)$ in $D^+$.\\
Suppose that $s_{11}(k)$ has $N$ simple zeroes in $D^+\bigcap\{{\rm Im}\ k>0\}$ denoted by $k_n, \ n=1,2,\ldots,N.$ Owing to the symmetries in (\ref{40}), there exists
$$s_{11}(k_n)=0\Longleftrightarrow s_{22}(-k_n^*)=0 \Longleftrightarrow s_{22}(k_n^*)=0\Longleftrightarrow s_{11}(-k_n)=0,$$
thus the discrete spectrum is the set
\begin{equation}
\{\pm k_n, \pm k_n^*\},
\end{equation}
which distribute in the $k$-plane as shown in Fig.2.
\begin{center}
\begin{tikzpicture}
\draw [<-](-2.8,0)--(-2.2,0);
\draw [-](-4,0)--(-2.8,0);
\draw [->](-1.4,0)--(-0.6,0);
\draw [->](-2.2,0)--(-1.4,0);

\draw [->](-2.2,-1.5)--(-2.2,-0.7);
\draw [-](-2.2,-0.7)--(-2.2,0);
\draw [->](-2.2,1.5)--(-2.2,0.7);
\draw [-](-2.2,0.7)--(-2.2,0);
\draw [->](-2.2,0.7)--(-2.2,1.5);

\node [thick] [above]  at (-1.5,0.8){\footnotesize $k_n$};
\node [thick] [above]  at (-1.4,0.7){\footnotesize $\centerdot$};

\node [thick] [above]  at (-1.5,-1.5){\footnotesize $k_n^*$};
\node [thick] [above]  at (-1.4,-1.1){\footnotesize $\centerdot$};

\node [thick] [above]  at (-2.9,0.8){\footnotesize $-k_n^*$};
\node [thick] [above]  at (-2.9,0.7){\footnotesize $\centerdot$};

\node [thick] [above]  at (-2.9,-1.4){\footnotesize $-k_n$};
\node [thick] [above]  at (-2.9,-1.1){\footnotesize $\centerdot$};
\node [thick] [above]  at (-2.2,1.5){\footnotesize ${\rm Im}k$};
\node [thick] [above]  at (-0.3,-0.2){\footnotesize ${\rm Re}k$};
\end{tikzpicture}
 \center{\footnotesize {\bf Fig.~2.} Discrete spectrum}
\end{center}

\indent Next we derive the residue conditions will be needed for the RH problem.
If $s_{11}(k)=0$ at $k=k_n$ the eigenfunctions $\varphi_{+,1}$ and $\varphi_{-,2}$ at $k=k_n$ must be proportional
\begin{equation}
\varphi_{+,1}(k_n)=b_n\varphi_{-,2}(k_n),\label{53}
\end{equation}
where $b_n$ is an arbitrary constant independent on $x,t$. Under the transformation (\ref{muvar}), there exists linear relation for $\mu_{\pm}$
\begin{equation}
\mu_{+,1}(k_n)=b_ne^{2i\theta(k_n)}\mu_{-,2}(k_n).
\end{equation}
Thus we get
\begin{equation}
{\rm Res}_{k=k_n}\left[ \frac{\mu_{+,1}}{s_{11}} \right]=\frac{\mu_{+,1}(k_n)}{s_{11}'(k_n)}=C_n e^{2i\theta(k_n)}\mu_{-,2}(k_n),
\end{equation}
where $C_n=\frac{b_n}{s_{11}'(k_n)}$.
As for $k=-k_n$, substituting  (\ref{42}) into (\ref{53}) leads to
\begin{equation}
\varphi_{+,1}(-k_n)=-b_n\varphi_{-,2}(-k_n),
\end{equation}
then applying the relation (\ref{muvar}) yileds
\begin{equation}
\mu_{+,1}(-k_n)=-b_ne^{2i\theta(k_n)}\mu_{-,2}(-k_n),
\end{equation}
thus we obtain
\begin{equation}
{\rm Res}_{k=-k_n}\left[ \frac{\mu_{+,1}}{s_{11}} \right]=\frac{\mu_{+,1}(-k_n)}{s_{11}^{'}(-k_n)}=-C_n e^{2i\theta(k_n)}\sigma_3\mu_{-,2}(k_n).
\end{equation}
\indent Similarly, the residue conditions at $k=\pm k_n^*$ are
\begin{align}
{\rm Res}_{k=k_n^*}\left[ \frac{\mu_{+,2}}{s_{22}} \right]&=\frac{\mu_{+,2}(k_n^*)}{s_{22}^{'}(k_n^*)}=\widetilde{C}_n e^{-2i\theta(k_n^*)}\mu_{-,1}(k_n^*),\\
{\rm Res}_{k=-k_n^*}\left[ \frac{\mu_{+,2}}{s_{22}} \right]&=\frac{\mu_{+,2}(-k_n^*)}{s_{22}^{'}(-k_n^*)}=\widetilde{C}_n e^{-2i\theta(k_n^*)}\sigma_3\mu_{-,1}(k_n^*),
\end{align}
where $\widetilde{C}_n=-C_n^*$.\\
\indent Recall the defination of $M^{\pm}$, there follows
\begin{equation}
\begin{split}
& {\rm Res}_{k=k_n^*}M^+=\left(0,\widetilde{C}_ne^{-2i\theta(k_n^*)}\mu_{-,1}(k_n^*)\right),\quad {\rm Res}_{k=-k_n^*}M^+=\left(0,\widetilde{C}_ne^{-2i\theta(k_n^*)}\sigma_3\mu_{-,1}(k_n^*)\right),\\
&{\rm Res}_{k=k_n}M^-=\left(C_ne^{2i\theta(k_n)}\mu_{-,2}(k_n),0\right),\quad {\rm Res}_{k=-k_n}M^-=\left(-C_ne^{2i\theta(k_n)}\sigma_3\mu_{-,2}(k_n),0\right).
\end{split}\label{58}
\end{equation}
\subsection{Reconstruction formula for the potential}
To solve the Riemann-Hilbert problem (\ref{RHP1}),one needs to regularize it by subtracting out the asymptotic behaviour and the pole contribution. Hence, we  rewrite Eq. (\ref{RHP1}) as
\begin{equation}
\begin{split}
&M^+-I-\frac{1+\beta}{k}e^{-i\nu\sigma_3}\widetilde{Q}_--\sum_{n=1}^N\frac{{\rm Res}_{k=k_n^*}M^+}{k-k_n^*}-\sum_{n=1}^N\frac{{\rm Res}_{k=k_n}M^-}{k-k_n}\\
-&\sum_{n=1}^N\frac{{\rm Res}_{k=-k_n^*}M^+}{k+k_n^*}-\sum_{n=1}^N\frac{{\rm Res}_{k=-k_n}M^-}{k+k_n}\\
=&M^--I-\frac{1+\beta}{k}e^{-i\nu\sigma_3}\widetilde{Q}_--\sum_{n=1}^N\frac{{\rm Res}_{k=k_n^*}M^+}{k-k_n^*}-\sum_{n=1}^N\frac{{\rm Res}_{k=k_n}M^-}{k-k_n}\\
-&\sum_{n=1}^N\frac{{\rm Res}_{k=-k_n^*}M^+}{k+k_n^*}-\sum_{n=1}^N\frac{{\rm Res}_{k=-k_n}M^-}{k+k_n}-M^-G,
\end{split}
\end{equation}
where $\widetilde{Q}_-=\left( \begin{array}{cc}
                        0  & -\frac{1}{q_-^*}\\
                        \frac{1}{q_-} &  0
                       \end{array}\right) $.
Then  the Plemelj's formula shows
\begin{equation}
\begin{split}
M(x,t,k)=&I+\frac{1+\beta}{k}e^{-i\nu\sigma_3}\widetilde{Q}_-+\sum_{n=1}^N\frac{{\rm Res}_{k=k_n^*}M^+}{k-k_n^*}+\sum_{n=1}^N\frac{{\rm Res}_{k=k_n}M^-}{k-k_n}\\
+&\sum_{n=1}^N\frac{{\rm Res}_{k=-k_n^*}M^+}{k+k_n^*}+\sum_{n=1}^N\frac{{\rm Res}_{k=-k_n}M^-}{k+k_n}+\frac{1}{2\pi i}\int_{\Sigma}\frac{M^-G(x,t,\xi)}{\xi-k}d\xi,
\end{split}\label{plemelj}
\end{equation}
and the $(1,2)$-element of $M$ is
\begin{equation}
\begin{split}
M_{12}=&\frac1k\bigg\{(1+\beta)e^{-i\nu\sigma_3}\widetilde{Q}_-+\sum_{n=1}^N   \big({\rm Res}_{k=k_n^*}M^++{\rm Res}_{k=-k_n^*}M^+\big)\\
&-\frac{1}{2\pi i}\int_{\Sigma}\frac{M^-G(x,t,\xi)}{\xi-k}d\xi  \bigg\}_{12}+O(1/k^2).
\end{split}
\label{1/k}
\end{equation}
\indent Comparing the $(1,2)$-element on the both sides of Eq. (\ref{21}) yields
\begin{equation}
q_x-i\beta q=2ie^{i\nu}J_{\pm,12}^{(1)}.\label{65}
\end{equation}
Recall the transformation (\ref{qu}), we know
$$u_xe^{i\tau(x,t)}=q_x-i\beta q,$$where $\tau(x,t)=\beta x+\alpha\beta^2q_0^2t.$
Thus we can write (\ref{65}) as
\begin{equation}
u_x=2ie^{i\nu-i\tau}\lim_{k\rightarrow\infty}\left( kJ_{\pm}\right)_{12}=2ie^{2i\nu-i\tau}\lim_{k\rightarrow\infty}\left( k\mu_{\pm}\right)_{12}=2ie^{2i\nu-i\tau}\left( M^-_1  \right)_{12},
\label{62}
\end{equation}
where
$$M^-=M^-_0+\frac{M^-_1}{k}+\cdots.$$
Substituting (\ref{1/k}) and (\ref{58}) into (\ref{62}), we  obtain the reconstruction formula for potential
\begin{equation}
\begin{split}
u_x=&2ie^{2i\nu-i\tau}\bigg\{ -\frac{(1+\beta)e^{-i\nu}}{q_-^*}+ 2\sum_{n=1}^N\widetilde{C}_ne^{-2i\theta(k_n^*)}\mu_{-,1,1}(k_n^*)\\&-\frac{1}{2\pi i}\int_{\Sigma} \big( M^-G\big)_{12}(\xi)d\xi \bigg\}.
\end{split}
\label{recon}
\end{equation}
\section{Reflectionless potentials}
\noindent Now we consider the potential $u(x,t)$ for which the reflection coefficient $\rho(k)$ vanishes identically, that is, $G=0$. In this case, Eq. (\ref{recon}) reads as
\begin{equation}
u_x=2ie^{-i\tau}\left(\sum_n 2e^{2i\nu}\widetilde{C}_ne^{-2i\theta(k_n^*)}\mu_{-,1,1}(k_n^*)-\frac{(1+\beta)e^{i\nu}}{q_-^*} \right).
\end{equation}
\indent To obtain the expression of the term $\mu_{-,1,1}(k_n^*)$, we consider  the first and second column of  (\ref{plemelj}) respectively under reflectionless case:
\begin{equation}
\begin{split}
&\mu_{-,2}(k_n)=\left(\begin{array}{c}
                -\frac{(1+\beta)e^{-i\nu}}{k_nq_-^*} \\
                1\end{array}\right)+\sum_{j=1}^{N}\frac{\widetilde C_je^{-2i\theta(k_j^*)}}{k_n-k_j^*}\mu_{-,1}(k_j^*)+\sum_{j=1}^{N}\frac{\widetilde C_je^{-2i\theta(k_j^*)}}{k_n+k_j^*}\sigma_3\mu_{-,1}(k_j^*),\\
&\mu_{-,1}(k_n^*)=\left(\begin{array}{c}
                  1\\
                  \frac{(1+\beta)e^{i\nu}}{k_n^*q_-} \end{array}\right)+\sum_{j=1}^{N}\frac{C_je^{2i\theta(k_j)}}{k_n^*-k_j}\mu_{-,2}(k_j)-\sum_{j=1}^{N}\frac{C_je^{2i\theta(k_j)}}{k_n^*+k_j}\sigma_3\mu_{-,2}(k_j),
\end{split}
\end{equation}
which can be further written as
\begin{equation}
\begin{split}
&\mu_{-,2}(k_n)=\left(\begin{array}{c}
                -\frac{(1+\beta)e^{-i\nu}}{k_nq_-^*} \\
                1\end{array}\right)+2\sum_{j=1}^{N}\frac{\widetilde C_je^{-2i\theta(k_j^*)}}{k_n^2-(k_j^{*})^2}K_1\mu_{-,1}(k_j^*),\\
&\mu_{-,1}(k_n^*)=\left(\begin{array}{c}
                  1\\
                  \frac{(1+\beta)e^{i\nu}}{k_n^*q_-} \end{array}\right)+2\sum_{j=1}^{N}\frac{C_je^{2i\theta(k_j)}}{(k_n^*)^2-k_j^2}K_2\mu_{-,2}(k_j),
\label{1soliton}
\end{split}
\end{equation}
where
\begin{equation}
K_1=\left(\begin{array}{cc}
     k_n & 0\\
     0   & k_j^*
     \end{array}\right),\quad K_2=\left(\begin{array}{cc}
     k_j & 0\\
     0   & k_n^*
     \end{array}\right).
\nonumber
\end{equation}
Define
$$ c_j(x,t,k)=\frac{C_j}{(k^*)^2-k_j^2} e^{2i\theta(x,t,k_j)},$$
whose 	conjugate  gives
$$c_j^*(k)=-\frac{\widetilde C_j}{k^2-(k_j^*)^2}e^{-2i\theta(k_j^*)}.$$
Then (\ref{1soliton}) reduces to
\begin{align}
\mu_{-,1,1}(k_n^*)&=1+2\sum_{j=1}^{N}k_jc_j(k_n)\mu_{-,1,2}(k_j),\label{mu11}\\
\mu_{-,1,2}(k_j)&=-\frac{(1+\beta)e^{-i\nu}}{k_jq_-^*} -2\sum_{m=1}^{N}k_j c_m^*(k_j)\mu_{-,1,1}(k_m^*),\label{mu12}
\end{align}
substituting (\ref{mu12}) into (\ref{mu11}) yields
\begin{equation}
\mu_{-,1,1}(k_n^*)=1-\frac{(2+2\beta)e^{-i\nu}}{q_-^*} \sum_{j=1}^{N}c_j(k_n) -4\sum_{j=1}^{N}\sum_{m=1}^{N}k_j^2c_j(k_n)c_m^*(k_j)\mu_{-,1,1}(k_m^*),\quad n=1,2,\ldots,N.
\label{mu}
\end{equation}

Introducing notations
$$X=(X_1,X_2,\ldots,X_{N})^t,\quad A=(A_{n,m}),\quad B=(B_1,B_2,\ldots,B_{N})^t$$
with components being
$$X_n=\mu_{-,1,1}(k_n^*),\quad A_{nm}=\sum_{j=1}^{N}4k_j^2c_j(k_n)c_m^*(k_j),\quad B_n=1-\frac{(2+2\beta)e^{-i\nu}}{q_-^*} \sum_{j=1}^{N}c_j(k_n),$$
then the system (\ref{mu})  can be written as matrix form
\begin{equation}
H X=B,
\label{system}
\end{equation}
where
$$H=I+A=(H_1,H_2,\ldots,H_{N}).$$

By standard  Crammer rule, the system (\ref{system}) is the solution of
\begin{equation}
\mu_{-,1,1}(k_n^*)=X_n=\frac{det\ H_n^{ext}}{det\ H}, \label{pop}
\end{equation}
where
$$H_n^{ext}=(H_1,\ldots,H_{n-1},B,\ldots,H_{N}).$$
Note that
$$u_xv_x=q_xr_x+i\beta q_x r-i\beta q r_x+\beta^2qr,$$
then Eq. (\ref{27}) reduces to
\begin{equation}
\nu=\int_{-\infty}^x(  \beta+\frac12u_xv_x)(x^{'},t)dx^{'},
\end{equation}
Therefore, we obtain a compact solution:
\begin{equation}
u_x=2ie^{-i\tau(x,t)+i\nu}\bigg(\frac{{\rm det}\ H^{aug}}{{\rm det}\ H}e^{i\nu}- \frac{1+\beta}{q_-^*}\bigg),
\label{82}
\end{equation}
where the augmented $(N+1)\times(N+1)$ matrix $H^{aug}$ is
\begin{equation}
H^{aug}=\left(\begin{array}{cc}
         0&Y^t\\
         B&H\end{array}\right),\quad Y=(Y_1,\ldots,Y_{N})^t,
\nonumber
\end{equation}
 and $Y_n=2\widetilde C_ne^{-2i\theta(k_n^*)}=- 2C_n^*e^{-2i\theta(k_n^*)}$.
\section{Trace formula and theta condition}
Define
\begin{equation}
\beta^-=s_{11}(k)\prod_{n=1}^N\frac{k^2-(k_n^*)^2}{k^2-k_n^2},\quad \beta^+=s_{22}(k)\prod_{n=1}^N\frac{k^2-k_n^2}{k^2-(k_n^*)^2},
\label{80}
\end{equation}
we see that they are analytic and no zeros in $D^-$ and $D^+$, respectively. Moreover, $\beta^+\beta^-=s_{11}(k)s_{22}(k).$
Note that  ${\rm det}\ S(k)=s_{11}s_{22}-s_{21}s_{12}=1$, this implies
$$\frac{1}{s_{11}s_{22}}=1-\rho(k)\widetilde{\rho}(k)=1+\rho(k)\rho^*(k^*),$$
thus
$$\beta^+\beta^-=s_{11}s_{22}=\frac{1}{1+\rho(k)\rho^*(k^*)},\quad k\in\Sigma.$$
Taking logarithms leads to
$$\log \beta^+-(-\log \beta^-)=-\log[1+\rho(k)\rho(k^*)],\quad k\in\Sigma,$$
then Applying Plemelj formula, we have
\begin{equation}
\log \beta^{\pm}=\mp\frac{1}{2\pi i}\int_{\Sigma}\frac{\log[1+\rho(s)\rho^*(s^*)]}{s-k}ds,\quad k\in D^{\pm}.
\end{equation}
Substituting into (\ref{80}), we obtain the trace formula
\begin{equation}
\begin{split}
s_{11}(k)&=\exp\bigg[ \frac{1}{2\pi i}\int_{\Sigma}\frac{\log[1+\rho(s)\rho^*(s^*)]}{s-k}ds  \bigg]\prod_{n=1}^N\frac{k^2-k_n^2}{k^2-(k_n^*)^2},\quad k\in D^-,\\
s_{22}(k)&=\exp\bigg[ -\frac{1}{2\pi i}\int_{\Sigma}\frac{\log[1+\rho(s)\rho^*(s^*)]}{s-k}ds  \bigg]\prod_{n=1}^N\frac{k^2-(k_n^*)^2}{k^2-k_n^2},\quad k\in D^+.
\end{split}
\end{equation}
Under reflectionless condition, they reduce to
\begin{equation}
s_{11}(k)=\prod_{n=1}^N\frac{k^2-k_n^2}{k^2-(k_n^*)^2},\quad k\in D^-;\quad s_{22}(k)=\prod_{n=1}^N\frac{k^2-(k_n^*)^2}{k^2-k_n^2},\quad k\in D^+.
\label{85}
\end{equation}
Taking limit as $k\rightarrow 0$ for (\ref{85}) leads to
\begin{equation}
\frac{q_-}{\beta q^+}=\exp\bigg[-\frac{i}{2\pi} \int_{\Sigma}\frac{\log[1+\rho(s)\rho^*(s^*)]}{s}ds  \bigg]\exp\bigg[4i\sum_{n=1}^Narg(k_n)\bigg], \quad k\in D^-,
\end{equation}
note that $\beta$ is a positive constant, then we obtain the theta condition
\begin{equation}
\arg\bigg( \frac{q_-}{ q_+}\bigg)=-\frac{1}{2\pi}\int_{\Sigma}\frac{\log[1+\rho(s)\rho^*(s^*)]}{s}ds+4\sum_{n=1}^N arg(k_n).
\end{equation}
under reflectionless condition, we have
\begin{equation}
\arg\bigg( \frac{q_-}{ q_+} \bigg)=4\sum_{n=1}^N \arg(k_n) .
\end{equation}
\section{One-soliton solution}
As an application of the formula (\ref{82})of N-soliton solution, we construct one-soliton solution for the FL equation, which corresponding to $N=1$.
Then Eq. (\ref{82}) becomes
\begin{equation}
u_x=2ie^{-i\tau+i\nu}\bigg(  -\frac{2C^*_1e^{-2i\theta(k_1^*)}e^{i\nu}}{1+4k_1^2|c_1(k_1)|^2}+\frac{ (4+4\beta)C^*_1e^{-2i\theta(k_1^*)}c_1(k_1)}{  q_-^*(1+4k_1^2|c_1(k_1)|^2)     }   -   \frac{1+\beta}{q_-^*} \bigg),
\end{equation}
where $k_1$ is an eigenvalue, $C_1$ is an arbitrary constant and
\begin{equation}
\begin{split}
\theta(k_1)&=\frac{k_1\lambda_1 }{\sqrt{\alpha}}x+\lambda_1\eta_1t,\quad \lambda_1=\sqrt{\alpha}\bigg( k_1+\frac{\beta}{2k_1} \bigg),\\
\eta_1&=\sqrt{\alpha}\bigg( k_1-\frac{\beta}{2k_1} \bigg),\quad c_1(k_1)=\frac{C_1}{(k_1^*)^2-k_1^2}e^{2i\theta(k_1)}.
\end{split}
\end{equation}

%
%

\end{document}